\magnification=\magstep1
\hsize 32 pc
\vsize 42 pc
\baselineskip = 24 true pt
\def\cl{\centerline}
\def\vs {\vskip .4 true cm}
\cl {\bf A MODEL OF COMPOSITE QUARKS AND LEPTONS}
\vs
\cl {\bf T. PRADHAN}
\cl {Institute of Physics, Bhubaneswar-751005, INDIA}
\vs
\cl {\bf ABSTRACT}

In continuation of our model of composite electron, its companion neutrinos,
photon and electro-weak gauge bosons as well as their super-partners based
on extended Lorentz  group $SU(3)\otimes SU(3)$ in (8+1) dimensional 
space-time, we show that the remaining leptons, quarks, lepto-quarks and
gluons as well as their respective super-partners can be
considered as composites of fundamental particles belonging to the $5\otimes 5$
dimensional representation of the further extended Lorentz group
$SU(5)\otimes SU(5)$ in (24+1) dimensional space-time. The fundamental particles constituting
the composite ones do not posses quantum numbers of the latter such as charge,
hypercharge etc. Since according to our estimate based on experimental limit
on the electron life-time, the mass of the fundamental particles is in the
Planckian regime, the composite particles would break up into their
constituents at these energies and lose their quantum numbers. 
\vfill
\eject
In an earlier communication [1], we showed that the electron, its
companion neutrino and the electro-weak gauge-bosons as well as their super-partners
are composites of fundamental particles belonging to the $3\otimes 3$ dimensional
representation of the extended Lorentz group $SU(3)\otimes SU(3)$ in (8+1)
dimensional space-time which are held together by forces provided by
gauge-particles belonging to the $8\otimes 8$ dimensional representation.
The fine structure constant of the gauge coupling was found to be unity
from the vanishing of the neutrino mass and an estimate of $m > 10^{22}$ GeV
for the mass of the fundamental particles was obtained from the
experimental limit $\tau > 10^{23}$ years for the life-time of the electron
indicating thereby that the composite nature of electron, photon etc will be
revealed at Planckian energies. In this communication we extend this model
so as to encompass all leptons, quarks, lepto-quarks, gluons and their 
super-partners in addition to particles covered in the previous model. It is
shown that this is accomplished through the extended $SU(5)\otimes SU(5)$ 
Lorentz group in (24+1) dimensional space-time.

The Dirac equations in spinor representation (symmetrical representation)
in (24+1) dimensional space-time reads:
$$\eqalign{ (p_0-{1\over 4}\Lambda_a p_a)\xi & = m\eta\cr
 (p_0+{1\over 4}\Lambda_a p_a)\eta & = m\xi\cr}\eqno{(1)}$$
a = 1,2,......24
where
$$\xi = \bigg ( \matrix{\xi_1\cr \xi_2\cr .\cr .\cr \xi_5\cr} \bigg) \ \eta = 
\bigg  (\matrix{\eta_1\cr\eta_2\cr .\cr .\cr \eta_5\cr}\bigg )$$
and $\Lambda_a$ are SU(5) matrices. This pair of equations split into the
four pairs of equations (2) and (3) in (8+1) dimensional space-times:
$$\eqalign{ (p_0-{1\over 2}\lambda_i p_i)\xi^{(\alpha)} & = m\eta^{(\alpha)}\cr
 (p_0+{1\over 2}\lambda_i p_i)\eta^{(\alpha)} & = m\xi^{(\alpha)}\cr}\eqno{(2)}$$
where
$$\xi^{\alpha} = \bigg ( \matrix{\xi_1\cr \xi_2\cr \xi_{\alpha}\cr} \bigg) \ \eta^{\alpha} = 
\bigg  (\matrix{\eta_1\cr\eta_2\cr \eta_{\alpha}\cr}\bigg )$$
$$\alpha = 3,4,5$$
$$ i = \cases {1,2,3,4,5,6,7,8 & for $\alpha$ = 3\cr\hfil&\hfil\cr
1,2,3,9,10,13,14,24 & for $\alpha$ = 4\cr\hfil&\hfil\cr
1 ,2,3,11,12,15,16,24 & for $\alpha$ = 5}$$
and 
$$\eqalign{ (p_0-{1\over 2}\lambda_i p_i)\xi^{(c)} & = m\eta^{(c)}\cr
 (p_0+{1\over 2}\lambda_i p_i)\eta^{(c)} & = m\xi^{(c)}\cr}\eqno{(3)}$$
where 
$$\xi^{(c)} = \bigg ( \matrix{\xi_4\cr \xi_5\cr \xi_3\cr} \bigg) \ \eta^{(c)} = 
\bigg  (\matrix{\eta_4\cr\eta_5\cr \eta_3\cr}\bigg )$$

$$i = 17,18,19,20,21,22,23,24$$

\noindent Each of these pairs of equations can be further reduced to ten pairs
of (3+1) space-time equations (4), (5), (6) and (7):
$$\eqalign{(p_0-\vec\sigma.\vec p) \xi_p & = m\eta_p\cr
p_0+\vec\sigma.\vec p) \eta_p & = m\xi_p\cr}\eqno{(4)}$$
where 
$$\xi_p = \bigg ( \matrix{\xi_1\cr \xi_2\cr} \bigg) \ \eta_p = 
\bigg  (\matrix{\eta_1\cr\eta_2\cr}\bigg )$$
$$ \vec p =  (p_1,p_2,p_3)$$
$$\eqalign{(p_0-\vec\sigma.\vec\pi^{(\alpha)})\xi^{(\alpha)}_{\pi} &  = m\eta^{(\alpha)}_{\pi}\cr
(p_0+\vec\sigma.\vec\pi^{(\alpha)} \eta^{(\alpha)}_{\pi} &  = m\xi^{(\alpha)}_{\pi}\cr}\eqno{(5)}$$
where 
$$\xi^{(\pi)}_{\alpha} = \bigg ( \matrix{\xi_1\cr \xi_{\alpha}\cr} \bigg) \ \alpha = 3,4,5$$
$$\vec\pi^{(3)} = (p_4,p_5,{p_8\over\sqrt 3}), \
\vec\pi^{(4)} = (p_9,p_{10},{p_{24}\over\sqrt{15}}), \
\vec\pi^{(5)} = (p_{15},p_{16},{p_{24}\over\sqrt 15})$$
$$\eqalign{(p_0-\vec\sigma.\vec \kappa^{(\alpha)} ) \xi_{\kappa}^{(\alpha)} = m\eta_{\kappa}^{(\alpha)}\cr
(p_0+\vec\sigma.\vec \kappa^{(\alpha)} ) \eta_{\kappa}^{(\alpha)} = m\xi_{\kappa}^{(\alpha)}\cr}\eqno{(6)}$$
where 
$$\xi^{(\alpha)}_{\kappa} = \bigg ( \matrix{\xi_2\cr \xi_{\alpha}\cr} \bigg) \ \alpha = 3,4,5$$
$$\vec\kappa^{(3)} = (p_6,p_7,{p_8\over\sqrt 3}), 
\vec\kappa^{(4)} = (p_{13},p_{14},{p_{24}\over\sqrt{15}}), \
\vec\kappa^{(5)} = (p_{11},p_{12},{p_{24}\over\sqrt 15})$$
$$\eqalign{(p_0-\vec\sigma.\vec p^{(r)} ) \xi_r = m\eta_r\cr
(p_0+\vec\sigma.\vec p^{(r)} ) \eta_r = m\xi_r\cr}\eqno{(7)}$$
where
$$\xi_r = \bigg ( \matrix{\xi_4\cr \xi_5\cr} \bigg) \ \eta_r = 
\bigg  (\matrix{\eta_4\cr\eta_5\cr}\bigg )$$
$$ \vec p_r = (p_{21},p_{22},p_{23})$$
and identical equations for $(\xi^{(b)},\eta^{(b)})$ and $(\xi^{(g)},\eta^{(g)})$
with
$$\xi_b = \bigg ( \matrix{\xi_4\cr \xi_3\cr} \bigg), \ \eta_b = 
\bigg  (\matrix{\eta_4\cr\eta_3\cr}\bigg ) , \
\xi_g = \bigg ( \matrix{\xi_5\cr \xi_3\cr} \bigg), \ \eta_g = 
\bigg  (\matrix{\eta_5\cr\eta_3\cr}\bigg ) $$
$$ \vec p_b (p_{19},p_{20},{p_{24}\over\sqrt {15}}),
\vec p_g (p_{17},p_{18},{p_{24}\over\sqrt {15}}) $$
Construction of composite particles can be made with the aid of 
scalars and spinors obtained from the ten fundamental fields
$$\xi_p = \bigg ( \matrix{\xi_1\cr \xi_2\cr} \bigg), \ \xi_{\pi}^{(\alpha)} = 
\bigg  (\matrix{\xi_1\cr\xi_{\alpha}\cr}\bigg ) , \
\xi_{\kappa}^{(\alpha)} = \bigg ( \matrix{\xi_2\cr \xi_{\alpha}\cr} \bigg) $$
$$\alpha = 3,4,5$$
$$\xi_r = \bigg ( \matrix{\xi_4\cr \xi_5\cr} \bigg), \ \xi_b = 
\bigg  (\matrix{\xi_4\cr\xi_3\cr}\bigg), \  \xi_g = 
\bigg  (\matrix{\xi_5\cr\xi_3\cr}\bigg)$$
the equations of motion of which are given in eqns(4), (5), (6) and (7).
The composition of various particles is as follows:

Leptons:

$$l^{(\alpha)} = \bigg ( \matrix{ e & \mu & \tau \cr \nu_e & \nu_{\mu} & \nu_{\tau}\cr}\bigg ) =
\xi_p\bigg ( \matrix{\xi_{\pi}^{(\alpha)^+}\cr \xi_{\kappa}^{(\alpha)^+}\cr}\bigg )\eqno{(8)}$$

Quarks :
$$q_i^{(\alpha)} = \bigg (\matrix{u_i & c_i & t_i\cr d_i & s_i & b_i\cr}\bigg ) = \Phi_{p_i}
\bigg ( \matrix{\xi_{\pi}^{(\alpha)^+}\cr \xi_{\kappa}^{(\alpha)^+}\cr}\bigg )\eqno{(9)}$$
where $\Phi_{p_i} = \xi_p\xi_i^+$ is a first rank spinor in the $(p_0,p_1,p_2,p_3)$ space.

Electro-weak Gauge Bosons :

$$ A = \Psi_p\phi_{\pi} \ \ Z = \Psi_p\phi_{\kappa} \ \ W = \Psi_p\phi_{\pi\kappa}\eqno{(10)}$$
where $\Psi_p = \xi_p\xi_p^+$ is a second rank spinor and
$$\phi_{\pi} = {1\over\sqrt 3}\sum_{\alpha}\xi^{(\alpha)^+}_{\pi} \xi^{(\alpha)}_{\pi}, \
\phi_{\kappa} = {1\over\sqrt 3}\sum_{\alpha}\xi^{(\alpha)^+}_{\kappa} \xi^{(\alpha)}_{\kappa}, \
\phi_{\pi\kappa} = {1\over\sqrt 3}\sum_{\alpha}\xi^{(\alpha)^+}_{\pi} \xi^{(\alpha)}_{\kappa}$$
are scalars.

Gluons :

$$g_{ij} = \Psi_p\phi_{ij}, \ i,j = r,b,g\eqno{(11)}$$
where $\phi_{ij} = \xi^+_i\xi_j$ is a scalar.

Lepto-quarks :
$$ X_i = \Psi_p\phi_{\pi}\xi_i, \ Y_i = \Psi_p\phi_{\kappa}\xi_i \eqno{(12)}$$
In all these constructions we have only the $\xi$ components of the particles concerned.
Their $\eta$ components can be obtained by parity reversal. Further, their spin states will
be obtained from those of $\xi_p$.

The super-partners of all the above particles can be constructed as follows:

Scalar Leptons :
$$ \tilde l^{(\alpha)} = \bigg ( \matrix{ \tilde e, & \tilde \mu & \tilde \tau \cr \tilde\nu_e & \tilde\nu_{\mu} & \tilde\nu_{\tau}\cr}\bigg ) =
\phi_p\bigg ( \matrix{\xi_{\pi}^{(\alpha)^+}\cr \xi_{\kappa}^{(\alpha)^+}\cr}\bigg )\eqno{(13)}$$
where $\phi_p = \xi^+_p\xi_p$ is a scalar.

Quarks :
$$\tilde q_i^{(\alpha)} = \bigg (\matrix{\tilde u_i & \tilde c_i & \tilde t_i\cr \tilde d_i & \tilde s_i & \tilde b_i\cr}\bigg ) = \phi_i
\bigg ( \matrix{\xi_{\pi}^{(\alpha)^+}\cr \xi_{\kappa}^{(\alpha)^+}\cr}\bigg )\eqno{(14)}$$
where $\phi_i = \phi_p\xi_i$ is a scalar.

Gluinos :
$$\tilde g_{ij} = \xi_p\phi_{ij}\eqno{(15)}$$

Gauginos:
$$\tilde A = \xi_p\phi_{\pi} \ \ \tilde Z = \xi_p\phi_{\kappa} \ \ \tilde W = \xi_p\phi_{\pi\kappa}\eqno{(16)}$$

As stated earlier, the gauge particles that provide the glue belong to the regular
representation of $SU(5)\otimes SU(5)$. The dynamics of these particles will be
similar to those of the $SU(3)\otimes SU(3)$ case discussed in our previous
communication, but more complex. We shall take up this problem in a later publication.

We conclude by remarking that the fundamental particles of our model do not posses
quantum numbers like charge, hypercharge etc which are attributes of the
composite particles. The latter will break up into the constituent fundamental particles
at Planckian energies and lose their quantum numbers. 

\vfill
\eject
\centerline {\bf References}
\vs
\item {1.} T. Pradhan, Institute of Physics Preprint IP/BBSR/97-46.
\vfill
\eject
\end